\documentclass[nojss]{jss}

\usepackage{orcidlink,thumbpdf,lmodern}

\usepackage{framed}

\usepackage{amsmath, amsthm}
\usepackage{bm, subcaption}
\usepackage{booktabs}

\author{Brian Buckley~\orcidlink{0000-0002-5478-5830}\\University College Dublin
   \And Adrian O'Hagan\\University College Dublin \And Marie Galligan\\University College Dublin}
\Plainauthor{Brian Buckley, Adrian O'Hagan, Marie Galligan}

\title{\pkg{VBphenoR}: A variational Bayes latent class approach for EHR-based patient phenotyping in \proglang{R}}
\Plaintitle{A variational Bayes latent class approach for EHR-based patient phenotyping in R}
\Shorttitle{A variational Bayes latent class approach for EHR-based patient phenotyping in R}

\Abstract{
The \pkg{VBphenoR} package for \proglang{R} provides a closed-form variational Bayes approach
to patient phenotyping using Electronic Health Records (EHR) data. We implement a variational Bayes
Gaussian Mixture Model (GMM) algorithm using closed-form coordinate ascent variational
inference (CAVI) to determine the patient phenotype latent class.  We then implement a variational
Bayes logistic regression, where we determine the probability of the phenotype in the supplied EHR
cohort, the shift in biomarkers for patients with the phenotype of interest versus a healthy population
and evaluate predictive performance of binary indicator clinical codes and medication codes. The
logistic model likelihood applies the latent class from the GMM step to inform the conditional.
}

\Plainkeywords{variational inference, Bayes, patient phenotype, electronic health records, EHR, R}

\Address{
  Brian Buckley\\
  School of Mathematics and Statistics\\
  University College Dublin\\
  Belfield, Dublin, D04 C1P1\\
  E-mail: \email{brian.buckley.1@ucdconnect.ie}\\
  \emph{and}\\
  Adrian O'Hagan\\
  School of Mathematics and Statistics\\
  University College Dublin\\
  Belfield, Dublin, D04 C1P1\\
  \emph{and}\\
  Marie Galligan\\
  School of Medicine\\
  University College Dublin\\
  Belfield, Dublin, D04 C1P1\\

}

\begin{document}



\section[Introduction: Patient Phenotyping using EHR data]{Introduction: Patient Phenotyping using EHR data} \label{sec:intro}

As regulatory agencies increasingly recognise real-world evidence as a complement to traditional clinical trial
data, interest has grown in applying Bayesian methods across both interventional and observational research (\cite{Boulanger+Carlin:2021}.
A central objective in many clinical investigations is the delineation of patient subgroups that exhibit comparable
disease-related characteristics (\cite{He+Belouali+Patricoski+Lehmann+Ball+Anagnostou+Kreimeyer+Botsis:2023}). Electronic Health Records (EHR)
have become an important resource for such phenotypic analyses (\cite{Hripcsak+Albers:2013}).

Bayesian approaches to patient phenotyping in clinical observational studies have been limited by the
computational challenges associated with applying the Markov Chain Monte Carlo (MCMC) approach to real-world
data. \cite{Hubbard+Huang+Harton+Oganisian+Choi+Utidjian+Eneli+Bailey+Chen:2019} proposed a Bayes latent class model that
could be used in a general context for observational studies that use EHR data. They consider the common clinical
context where gold-standard phenotype information, such as genetic and laboratory data, is not fully available.
A general model of this form has high potential applicability for use in clinical decision support across
disease areas for both primary and secondary clinical databases.

Latent Class Analysis (LCA) is widely used when we want to identify patient phenotypes or subgroups
given multivariate data (\cite{Lanza+Rhoades:2013}).  A challenge in clinical LCA is the prevalence of
mixed data, where we may have combinations of continuous, nominal, ordinal and count data.  Bayesian
approaches to LCA may better account for this data complexity. The Bayesian approach serves as a
unifying paradigm between rule-based phenotyping approaches, which are predominantly informed by domain
expertise and clinical judgment (\cite{Cuker+Buckley+Mousseau+Barve+Haenig+Bussel:2023}), and data-driven
methods, which rely exclusively on empirical data and typically lack mechanisms for incorporating prior
or expert knowledge.

We previously implemented a patient phenotyping latent class model based on that proposed by
\cite{Hubbard+Huang+Harton+Oganisian+Choi+Utidjian+Eneli+Bailey+Chen:2019}, but adopted a variational Bayes
approach, as the Markov Chain Monte Carlo (MCMC) method was not computationally scalable for realistic
electronic health record (EHR) data (\cite{Buckley+O'Hagan+Galligan:2024}).  We used automatic
differentiation variational inference (ADVI) from \cite{Kucukelbir+Tran+Ranganath+Gelman+Blei:2017}.
We tested implementations of ADVI available in the \pkg{rstan} package (\cite{Carpenter+Gelman+Hoffman+Lee+Goodrich+Betancourt+Brubaker+Guo+Li+Riddell:2017})
and \pkg{PyMC3} \proglang{Python} library (\cite{Salvatier+Wiecki+Fonnesbeck:2016}). Although this
approach delivered reasonable results, it required significantly complex technical
tuning and multiple trial-and-error iterations. We found automatic VB methods as
implemented in \pkg{rstan} VB and \pkg{PyMC3} are complex to configure and are very sensitive to model
definition and algorithm hyperparameters such as choice of gradient optimiser. We found closed-form mean-field
VB performed well in our Pima Indians case study using the \proglang{R} packages \pkg{varbvs} (\cite{Carbonetto+Stephens:2012})
and \pkg{sparsevb} (\cite{Clara+Szabo+Ray:2025}).

We now propose a closed-form approach, implemented in the \proglang{R} package \pkg{VBphenoR},
based on the theory defined in \cite{Bishop+Nasser:2006}, \cite{Nakajima+Watanabe+Sugiyama:2019}
and the review of \cite{Blei+Kucukelbir+McAuliffe:2017}. We implement a variational
Bayes Gaussian Mixture Model (GMM) algorithm using the closed-form coordinate ascent
variational inference (CAVI) approach to determine the phenotype latent class.  We then
implement a variational Bayes logistic regression based on \cite{Durante+Rigon:2019},
where we determine the probability of the phenotype in the supplied cohort using the results
from the VB GMM classified into disease and non-disease classes. The VB logistic regression
uses informative priors on the coefficients to determine the shift in biomarkers for patients
with the phenotype of interest versus a normal population along with predictive performance
of binary indicator clinical codes and medication codes. The logit model likelihood uses the
latent class from the GMM step to inform the conditional (see Section~\ref{sec:algorithm}).

While the \proglang{R} environment for statistical computing features several packages for patient
phenotyping using EHR data, many are concerned with natural language processing approaches, such
as the use of International Classification of Diseases (ICD) codes and narrative data (\pkg{MAP}, \cite{Liao+Sun+Cai+Link+Hong+Huang+Huffman+Gronsbell+Zhang+Ho:2019}) and
latent Dirichlet allocation (\pkg{sureLDA}, \cite{Ahuja+Zhou+He+Sun+Castro+Gainer+Murphy+Hong+Cai:2020}).
Another \proglang{R} package, \pkg{PheVis} (\cite{Ferte+Cossin+Schaeverbeke+Barnetche+Jouhet+Hejblum:2021}), which
extends \pkg{PheNorm} (\cite{Yu+Ma+Gronsbell+Cai+Ananthakrishnan+Gainer+Churchill+Szolovits+Murphy+Kohane:2018}),
focuses on the probability of disease condition for patient visits using a machine learning approach. No dedicated
\proglang{R} package for performing patient phenotyping using EHR data within a Bayesian paradigm yet exists.

The rest of this paper is organised as follows: algorithm specification for patient
phenotyping in a Bayesian setting is briefy outlined in Section~\ref{sec:algorithm}.
Also included in this section is a general example of the package applied to its contained
dataset, the Sickle Cell Disease data. Details of the algorithm is provided in
Section~\ref{sec:details}. The effects of informative priors on the posterior
results are also discussed in detail through an application to the widely
available \code{faithful} dataset (\cite{Azzalini+Bowman:1990}) in Section~\ref{sec:informative prior GMM}.
This section has important guidance for choosing informative priors that generate clinically credible
results. For the variational GMM, we also study the effect of initialiser in
Section~\ref{sec:init}, an important consideration for unbalanced data, which is a common
feature of clinical datasets. A summary of the package, as well as potential future extensions
to the package are briefly discussed in Section~\ref{sec:summary}.

All computations and graphics in this paper have been carried out with \pkg{VBphenoR}
version 1.1. The latest version of the package should always be obtained using the
Comprehensive R Archive Network at \url{http://CRAN.R-project.org/package=VBphenoR}.



\section{Algorithm implementation} \label{sec:algorithm}

The EHR patient phenotyping model specification used in \cite{Buckley+O'Hagan+Galligan:2024}
associates the latent phenotype with patient characteristics and clinical history as
described in Table~\ref{table:modelSpec}.  $X_i$ represents $M$ patient covariates, such as
demographics ($X_i = X_{i1},...,X_{iM}$).  The parameter $\beta^D$ associates the latent phenotype to
patient characteristics, $\eta_i$ is a patient-specific random effect parameter and
parameters $\beta^R,\beta^Y,\beta^W$ and $\beta^P$ associate the latent phenotype and patient characteristics
to biomarker availability, biomarker values where available, clinical codes, and medications respectively.
The mean biomarker values are shifted by a regression quantity $\beta^Y_{j,M+1}$ for patients with the
phenotype compared to those without.

\begin{center}
\captionof{table}{Model specification for Bayesian latent variable model for EHR-derived phenotypes for patient $i$ (from \cite{Hubbard+Huang+Harton+Oganisian+Choi+Utidjian+Eneli+Bailey+Chen:2019}).}
\vspace{5mm}
\scalebox{0.80}{
\begin{tabular}{lllll}
\toprule
                                        & \multicolumn{1}{c}{\bfseries Variable} & \multicolumn{1}{c}{\bfseries Model} & \multicolumn{1}{c}{\bfseries Priors}   \\
\midrule
                                        &                       &                                                                &            \\
\textbf{Latent Phenotype}               & $D_i$                 & $D_i \sim $Bern$(g(\bm{X}_i\bm{\beta}^D + \eta_i))$            & $\beta^D \sim $MVN$(0, \Sigma_D);$ \\
                                        &                       &                                                                & $\eta_i \sim $Unif$(a,b)$              \\
                                        &                       &                                                                &            \\
\textbf{Availability of Biomarkers}     & $R_{ij}, j=1,...,J$   & $R_{ij} \sim $Bern$(g((1,\bm{X}_i,D_i)\bm{\beta}^R_j))$        & $\beta^R_j \sim $MVN$(\mu_R, \Sigma_R)$ \\
                                        &                       &                                                                &            \\
\textbf{Biomarkers}                     & $Y_{ij}, j=1,...,J$   & $Y_{ij} \sim $N$(g((1,\bm{X}_i,D_i)\bm{\beta}^Y_j, \tau^2_j))$ & $\beta^Y_j \sim $MVN$(\mu_Y, \Sigma_Y);$ \\
                                        &                       &                                                                & $\tau^2_j \sim $InvGamma$(c,d)$  \\
                                        &                       &                                                                &            \\
\textbf{Clinical Codes}                 & $W_{ik}, k=1,...,K$   & $W_{ik} \sim $Bern$(g((1,\bm{X}_i,D_i)\bm{\beta}^W_k))$        & $\beta^W_k \sim $MVN$(\mu_W, \Sigma_W)$ \\
                                        &                       &                                                                &            \\
\textbf{Prescription Medications}       & $P_{il}, l=1,...,L$   & $P_{il} \sim $Bern$(g((1,\bm{X}_i,D_i)\bm{\beta}^P_l))$        & $\beta^P_l \sim $MVN$(\mu_P, \Sigma_P)$ \\
                                        &                       &                                                                &            \\
\bottomrule
                                        &                       &                                                                &            \\
                                        &                       &       & \multicolumn{1}{c}{$\bm{ g(\bm{\cdot}) = exp(\bm{\cdot})/(1 + exp(\bm{\cdot})) }$} \\
\end{tabular}
}
\label{table:modelSpec}
\end{center}
\newpage
The likelihood for patient $i$ is given by:

\begin{align} \label{eq:likelihood}
\mathcal{L}(\eta_i,\beta^D,\beta^R,\beta^Y,\beta^W,\beta^P,\tau^2|X_i) &= \sum_{d=0,1}{P(D_i=d|\eta_i,\beta^D,X_i)}\\
& \prod_{j=1}^{J}{f(R_{ij}|D_i=d,X_i,\beta_j^R)f(Y_{ij}|D_i=d,X_i,\beta_j^Y,\tau_j^2)^{R_{ij}}}\nonumber \\
& \prod_{k=1}^{K}{f(W_{ik}|D_i=d,X_i,\beta_k^W)} \prod_{l=1}^{L}{f(P_{il}|D_i=d,X_i,\beta_i^P)}\nonumber \\ \nonumber
\end{align}

The likelihood for patient $i$ shows how biomarker availability affects the inclusion of the
likelihood contribution from $Y_{ij}$.  Biomarker data is considered gold-standard patient phenotype information.

The implementation of \pkg{VBphenoR} performs the following steps:

\begin{enumerate}
  \item Run a variational Gaussian Mixture Model using EHR-derived patient characteristics
    to discover the latent variable $D_i$ indicating the phenotype of interest for
    the $i^{th}$ patient. Patient characteristics can be any patient variables typically
    found in EHR data e.g.
    \begin{itemize}
      \item Gender
      \item Age
      \item Ethnicity (for disease conditions with genetic ethnicity-related increased risk)
      \item Physical e.g. BMI for Type 2 Diabetes
    \end{itemize}
  \item Run a variational regression model using the phenotype latent variable $D_i$ derived
    in step 1 to determine the shift in biomarker levels from the prior settings for patients
    with the phenotype versus those without. Appropriately informative values are used to set
    the biomarker prior, e.g. assuming a healthy value (see Table~\ref{table:modelSpec}).
  \item Run a variational logistic regression model using the latent variable ${D_i}$ derived
    in step 1 as an indicator for the phenotype of interest to determine the sensitivity and
    specificity of binary indicators for clinical codes, medications and availability of
    biomarkers (since biomarker laboratory tests will include a level of missingness that can
    vary by disease condition and can be random or systematic). In this framework, the sensitivity
    and specificity of binary indicators based on clinical codes, medication use, or biomarker
    presence can be expressed through combinations of the regression coefficients. For example,
    in a model that excludes patient-level covariates, the sensitivity of the $k^{th}$ clinical
    code is represented by expit($\beta^W_{k0} + \beta^W_{k1}$), whereas its specificity is
    calculated as 1 - expit($\beta^W_{k0}$). Here, expit($x$) denotes the logistic transformation
    exp($x$)/(1 + exp($x$)).
\end{enumerate}

\subsection{Patient phenotype example for rare Sickle Cell Disease}

The Sickle Cell Disease (SCD) data available in this package is a transformed version
of the SCD data from \cite{Al-Dhamari+Abu-Attieh+Prasser:2024}. We have retained a
subset of the data columns that are relevant to our model and transformed
the data into a representative cohort by retaining an expected prevalence of SCD (0.3\%),
with the rest converted to non-SCD patients by distributing the biomarker values around
a healthy value.

The following example illustrates the basic implementation.  We use the package internal SCD
data to find the rare SCD phenotype. As SCD is extremely rare, we use \code{DBSCAN}
(\pkg{dbscan}, \cite{Hahsler+Piekenbrock+Doran:2019})
to initialise the VB GMM. We also use an informative prior for the GMM mixing coefficient and
stop iterations when the ELBO starts to reverse so that we stop when the minor (SCD)
component is reached.  Section~\ref{sec:details} provides a detailed account of the
informative priors.  Section~\ref{sec:init} provides explanations of initialiser parameters.
In this basic example we are using a low value for the GMM $\alpha$ hyperparameter as we know the
data separates into two clusters so we want a prior for $\alpha$ that allows for the
posterior to be influenced more by the data.  For the VB logistic step, we use the mean
of the data for patient characteristics, age and highrisk ethnicity as well as the mean of
the biomarkers because these data are 99.7\% non-SCD patients. In this example we do not have
missing data.  CBC is the Complete Blood Count test (usually in g/dL) and RC is the Reticulocyte Count (\%).
Both are common laboratory biomarket diagnostic tests for patients with Anaemia
conditions.  For CBC, the normal range is approximately 12 g/dL and for RC it is between
0.5\% and 2.5\%.  Patients with SCD have reduced CBC and elevated RC.

\begin{CodeChunk}
\begin{CodeInput}
R> library(data.table)
R>
R> data(scd_cohort)
R> X1 <- scd_cohort[,.(CBC,RC)]
R> initParams <- c(0.15, 5)
R> names(initParams) <- c('eps','minPts')
R> X1 <- t(X1)
R>
R> prior_gmm <- list(
R>   alpha = 0.001
R> )
R>
R> prior_logit <- list(mu=c(1,
R>                    mean(scd_cohort$age),
R>                    mean(scd_cohort$highrisk),
R>                    mean(scd_cohort$CBC),
R>                    mean(scd_cohort$RC)),
R>               Sigma=diag(1,5))
R>
R> X2 <- scd_cohort[,.(age,highrisk,CBC,RC)]
R> X2[,age:=as.numeric(age)]
R> X2[,highrisk:=as.numeric(highrisk)]
R> X2[,Intercept:=1]
R> setcolorder(X2, c("Intercept","age","highrisk","CBC","RC"))
R>
R> biomarkers <- c('CBC', 'RC')
R> set.seed(123)
R>
R> pheno_result <- run_Model(biomarkers,
R>                         gmm_X=X1, gmm_init="dbscan",
R>                         gmm_initParams=initParams,
R>                         gmm_maxiters=20, gmm_prior=prior_gmm,
R>                         gmm_stopIfELBOReverse=TRUE,
R>                         logit_X=X2, logit_prior=prior_logit
R> )
R>
R> print(pheno_result$biomarker_shift)
\end{CodeInput}
\end{CodeChunk}

An important indicator in the model results is the shift in biomarker regression
prior for the phenotype of interest.  In the example we get:

\begin{table}[h]
\begin{tabular}{ll}
CBC\_shift & RC\_shift \\
\hfill7.93       & \hfill3.67
\end{tabular}
\end{table}

Our results indicate the patients with latent SCD in our cohort display both
biomarker levels indicative of SCD.



\section{Algorithm details} \label{sec:details}

\subsection{Variational Gaussian Mixture Model} \label{sec:Variational Gaussian Mixture Model}

We employ a variational Gaussian Mixture Model (GMM) to detect the latent class
for patients with the disease condition based on patient characteristics. The full
derivation for the GMM can be found in \cite{Bishop+Nasser:2006}. Here, we explain
how the posterior is affected by informative priors.  The derivation in \cite{Bishop+Nasser:2006}
uses conjugate priors to simplify the analysis so we implement the same conjugacy
in this package.

The conjugate prior for the GMM mixing coefficients, $\pi$, is a Dirichlet
distribution with prior hyperparameter $\alpha$.

\begin{equation} \label{eq:pi}
q(\bf{\pi}) = \textit{Dir}( \bf{\pi}|\alpha)
\end{equation}

The conjugate prior governing the unknown mean and precision of each GMM multivariate
Gaussian component is an independent Normal-Wishart prior.

\begin{equation} \label{eq:wishart}
q(\bf{\mu}, \bf{\Lambda}) = \mathcal{N}(\bf{\mu}|\bf{m}, (\lambda\bf{\Lambda})^{-1})\mathcal{W}(\bf{\Lambda}|\bf{W}, \nu)
\end{equation}

Where

\begin{itemize}
  \item $\lambda$ governs the proportionality of the precision, $\Lambda$ and influences
  the variance of $\mu$.
  \item $W$ is a symmetric, positive definite matrix with dimensions given by the
number of mixture components and governs the unknown precision of each GMM multivariate
Gaussian component.
  \item $m$ is a prior hyperparameter governing the mean vector for the Normal part and shrinks
  $\mu$ towards it. This is usually defaulted to the row means of the data.  A different
  value can be used in rare cases where there is some specific prior clinical evidence
  that the cluster means should fall in a specific region away from the data mean.
  \item $\nu$, the degrees of freedom, ensures the Wishart $\Gamma$ function is
well-defined. $\nu$ must be greater than $D-1$ (where $D$ is the dimensionality
of the data) to ensure the distribution is proper and the mean of the
Wishart ($\nu W^{-1}$) exists. This is usually defaulted to $D+1$ as it is the
smallest value that ensures the Wishart distribution is proper and the expected
value of the precision matrix exists. This value also results in a vague prior (one
that is weakly informative). It could be increased if there is a desire to avoid
overly flexible or noisy covariance estimates.
\end{itemize}

\subsection{Effect of informative priors on the Variational Gaussian Mixture Model} \label{sec:informative prior GMM}

Informative priors play an important role in the variational model. In this section
we briefly outline the prior effects on the VB GMM posterior estimates using the \code{faithful} data.
For a full explanation see \cite{Bishop+Nasser:2006}.  We use the kmeans \code{init} option for
illustration as the \code{faithful} data is simple to model.  For complex data, such as
our Sickle Cell Disease data, where the disease-positive class is extremely
unbalanced (0.3\%) and covered by a very noisy tail from the negative class, we
use \code{DBSCAN} instead (\cite{Mondragon+Lara+Eleuterio+Gutirrez+Lopez:2023}). \code{DBSCAN}
can perform well with non-spherical data and real-world datasets with noise.

\subsubsection{Prior hyperparameter alpha for the mixing Dirichlet distribution}

\begin{figure}[t!]
\centering
\begin{subfigure}{0.31\textwidth}
\includegraphics[width=\linewidth]{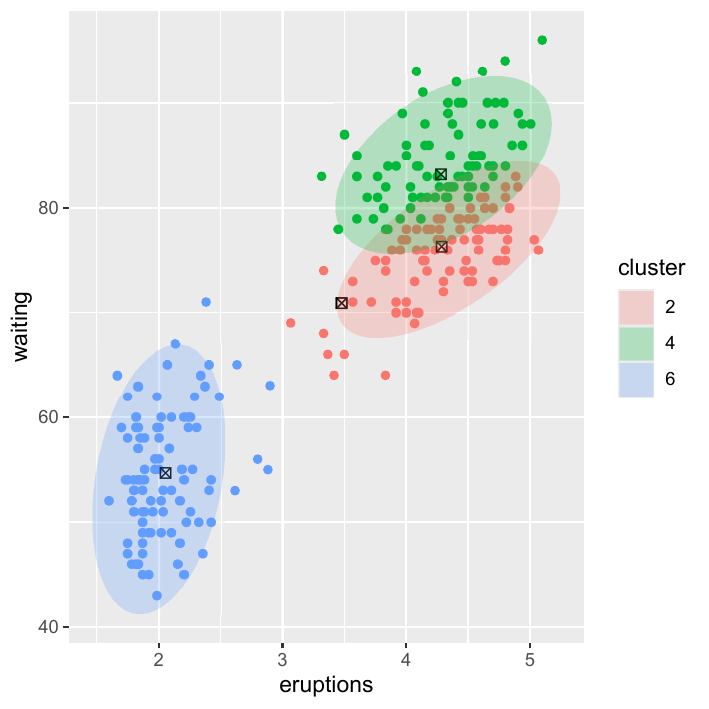}
\caption{$\alpha$ = 1, 271} \label{fig:1a}
\end{subfigure}
\begin{subfigure}{0.31\textwidth}
\includegraphics[width=\linewidth]{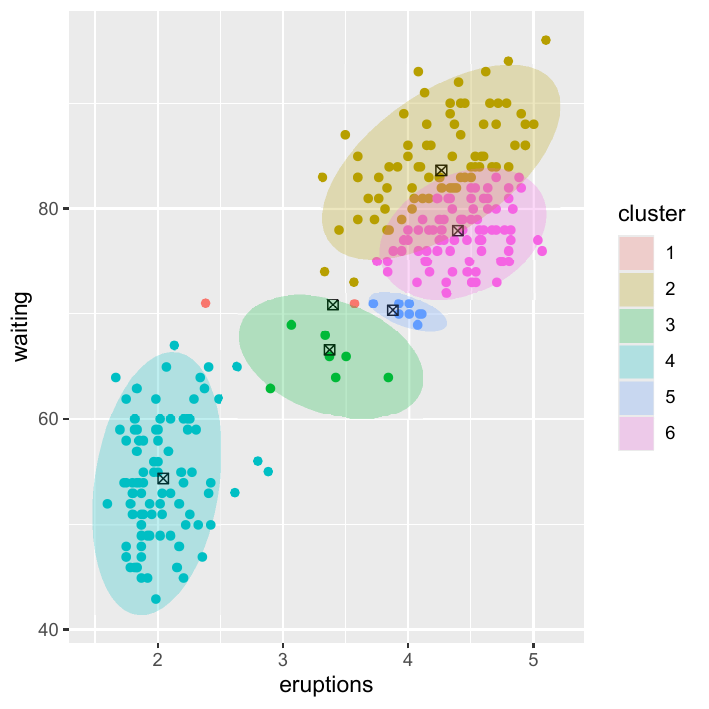}
\caption{$\alpha$ = 30, 237} \label{fig:1b}
\end{subfigure}
\begin{subfigure}{0.31\textwidth}
\includegraphics[width=\linewidth]{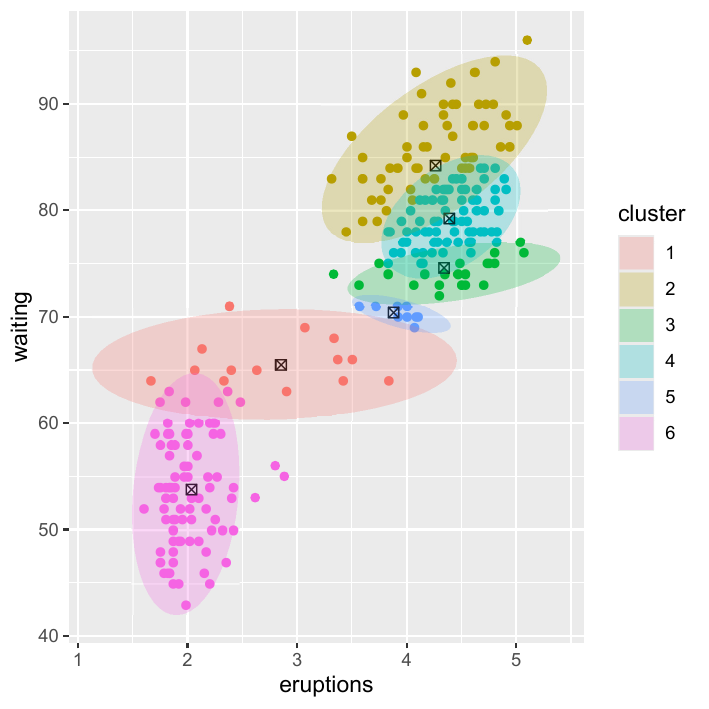}
\caption{$\alpha$ = 70, 202} \label{fig:1c}
\end{subfigure}
\caption{\label{fig:alpha1} Posterior clustering with three $\alpha$ prior hyperparameter
settings. (a) low $\alpha$ setting;  (b) medium $\alpha$ setting and (c) high $\alpha$ setting}
\end{figure}

The prior for the GMM mixing coefficients, $\pi$, affects the expectation such that
if $\alpha \longrightarrow 0$ the posterior distribution will be influenced
primarily by the data and as $\alpha \longrightarrow \inf$ then the prior will have
increased influence on the posterior.  To illustrate, in Figure~\ref{fig:alpha1}
and in the associated code, we use the base \proglang{R} \pkg{faithful} data with
three different values for $\alpha$ (low, medium and high) and set $k=6$ for the
\code{vb_gmm_cavi()} function. In this simplest case, we use the same $\alpha$ value
for each component. As alpha increases, the number of cluster components clearly
approaches the desired k. However, a lower alpha tends to produce a more intuitive
number of components for these data. In a realistic clinical setting, prior clinical
knowledge can be used to guide the selection of alpha, allowing us to balance the
number of components in a way that provides meaningful clinical insight while
minimising the risk of bias introduced by setting alpha too high.

\begin{figure}[t!]
\centering
\begin{subfigure}{0.31\textwidth}
\includegraphics[width=\linewidth]{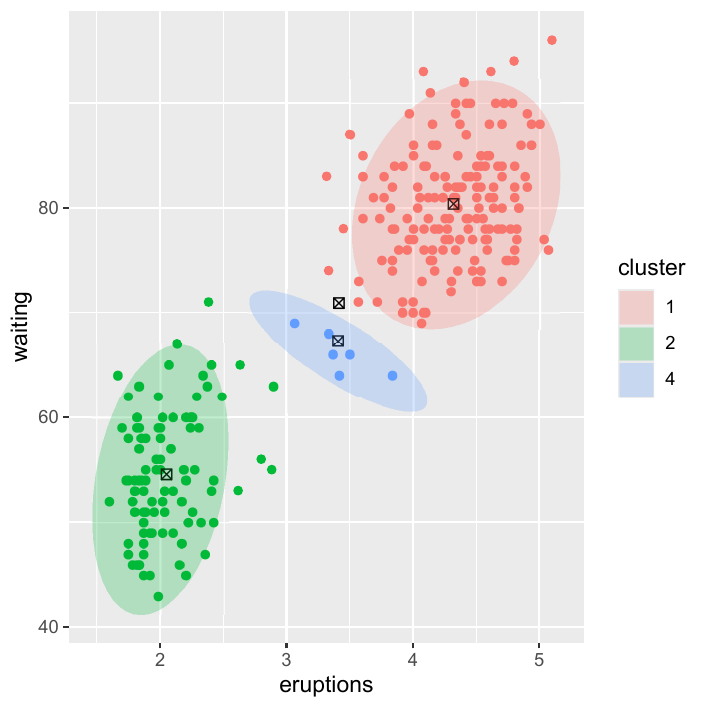}
\caption{$\alpha$ = 1,1,1,1} \label{fig:2a}
\end{subfigure}
\begin{subfigure}{0.31\textwidth}
\includegraphics[width=\linewidth]{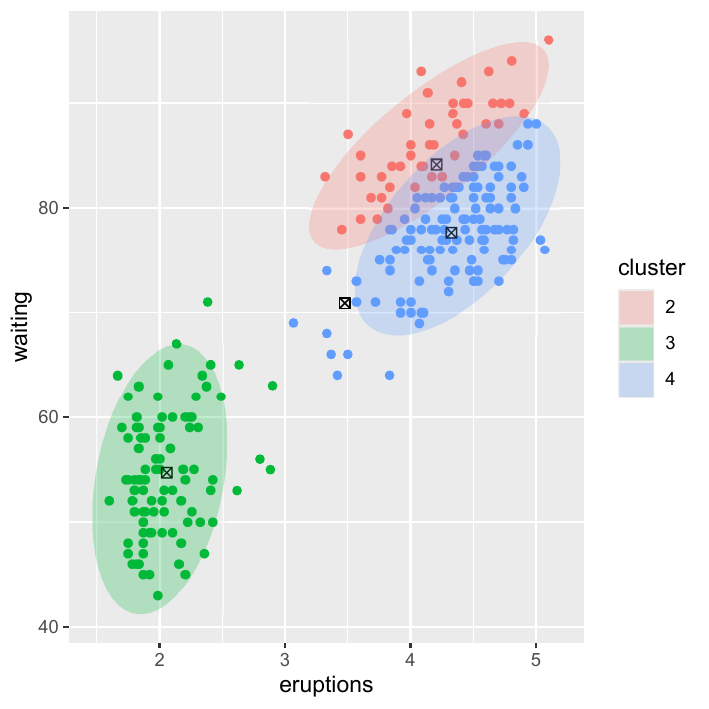}
\caption{$\alpha$ = 1,92,183,183} \label{fig:2b}
\end{subfigure}
\begin{subfigure}{0.31\textwidth}
\includegraphics[width=\linewidth]{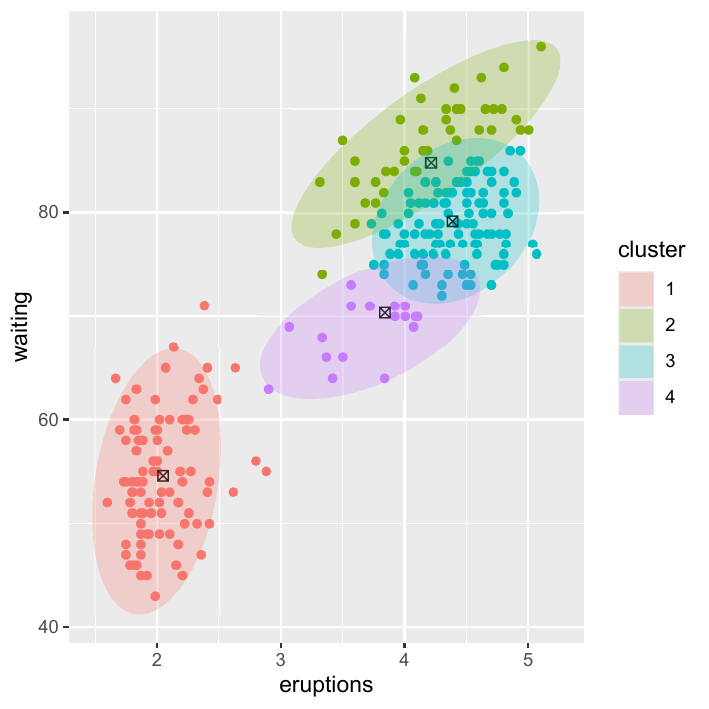}
\caption{$\alpha$ = 183,92,198,50} \label{fig:2c}
\end{subfigure}
\caption{\label{fig:alpha2} Posterior clustering with three different $\alpha$ vector
prior hyperparameter settings. (a) equal $\alpha$ vectors;  (b) two equal and two
different $\alpha$ vectors and (c) different $\alpha$ vectors.}
\end{figure}

If we use different $\alpha$ values for each cluster we can further fine-tune the
resulting posterior.  In Figure~\ref{fig:alpha2}, we illustrate three different settings where
the $\alpha$ per cluster is equal and set low (a), equal and set high (c)
and different per cluster component (b).  In this example we set $k=4$.  We use the
\code{faithful} data to show the effect of VB GMM Priors, stopping on ELBO reverse
parameter or delta threshold reached.

The code for Figure~\ref{fig:alpha2} is shown below.
(note the \code{do_prior_plots()} function is detailed in Appendix~\ref{app:plots}).

\newpage
\begin{CodeChunk}
\begin{CodeInput}
R>   gen_path <- tempdir()
R>   data("faithful")
R>   X <- faithful
R>   P <- ncol(X)
R>   k <- 4
R>
R>   alpha_grid <- data.frame(c1=c(1,1,1,1),
R>                            c2=c(1,92,183,183),
R>                            c3=c(183,92,198,50))
R>   init <- "kmeans"
R>
R>   for (i in 1:ncol(alpha_grid)) {
R>     prior <- list(
R>       alpha = as.integer(alpha_grid[,i])
R>     )
R>
R>     gmm_result <- vb_gmm_cavi(X=X, k=k, prior=prior, delta=1e-8, init=init,
R>                               verbose=FALSE, logDiagnostics=FALSE)
R>     do_prior_plots(i, gmm_result, "alpha", alpha_grid, gen_path)
R>   }
\end{CodeInput}
\end{CodeChunk}

\subsubsection{Prior hyperparameter lambda for the Normal-Wishart distribution.}

\begin{figure}[t!]
\centering
\begin{subfigure}{0.41\textwidth}
\includegraphics[width=\linewidth]{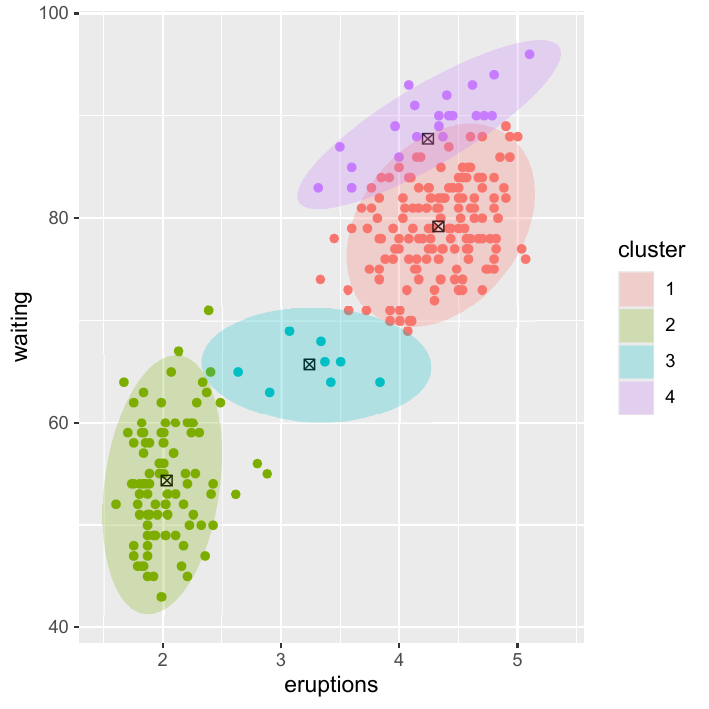}
\caption{$\lambda$ = 0.1} \label{fig:3a}
\end{subfigure}
\begin{subfigure}{0.41\textwidth}
\includegraphics[width=\linewidth]{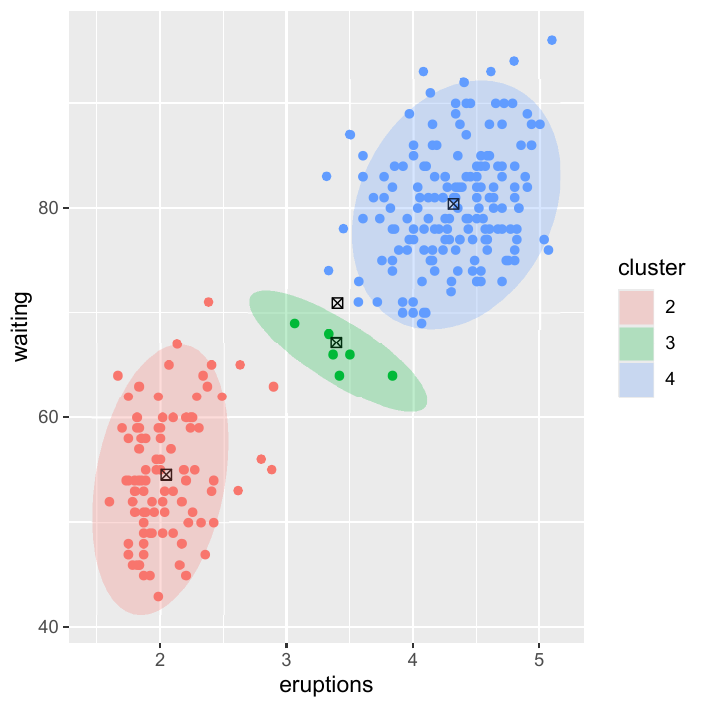}
\caption{$\lambda$ = 0.9} \label{fig:3b}
\end{subfigure}
\caption{\label{fig:lambda} Posterior clustering with two different lambda vector prior
hyperparameter settings. (a) low $\lambda$ vectors; (b) high $\lambda$ vectors.}
\end{figure}

The $\lambda$ prior determines the strength by which the prior mean, $m$, attracts cluster means
towards a chosen centre. Figure~\ref{fig:lambda} shows the effect of a very low value for
$\lambda$ and a very high value for $\lambda$.  Low values for $\lambda$ lead to weaker
distributions around $m$. In this case the posterior means, $\mu_k$, will be driven mainly
by the data assigned to each cluster.  On the contrary, high $\lambda$ values encode a strong
belief that component means are close to $m$.  The \proglang{R} code below shows that we set
a $\lambda$ prior for each cluster, in this case four.

\newpage
\begin{CodeChunk}
\begin{CodeInput}
R>   gen_path <- tempdir()
R>   lambda_grid <- data.frame(c1=c(0.1,0.1,0.1,0.1),
R>                             c2=c(0.9,0.9,0.9,0.9))
R>   init <- "kmeans"
R>   k <- 4
R>
R>   for (i in 1:ncol(lambda_grid)) {
R>     prior <- list(
R>       beta = as.numeric(lambda_grid[,i])
R>     )
R>
R>     gmm_result <- vb_gmm_cavi(X=X, k=k, prior=prior, delta=1e-8, init=init,
R>                               verbose=FALSE, logDiagnostics=FALSE)
R>     do_prior_plots(i, gmm_result, "lambda", lambda_grid, gen_path)
R>   }
\end{CodeInput}
\end{CodeChunk}

\subsubsection{Prior hyperparameter W for the scale matrix for the inverse Wishart.}

\begin{figure}[t!]
\centering
\begin{subfigure}{0.41\textwidth}
\includegraphics[width=\linewidth]{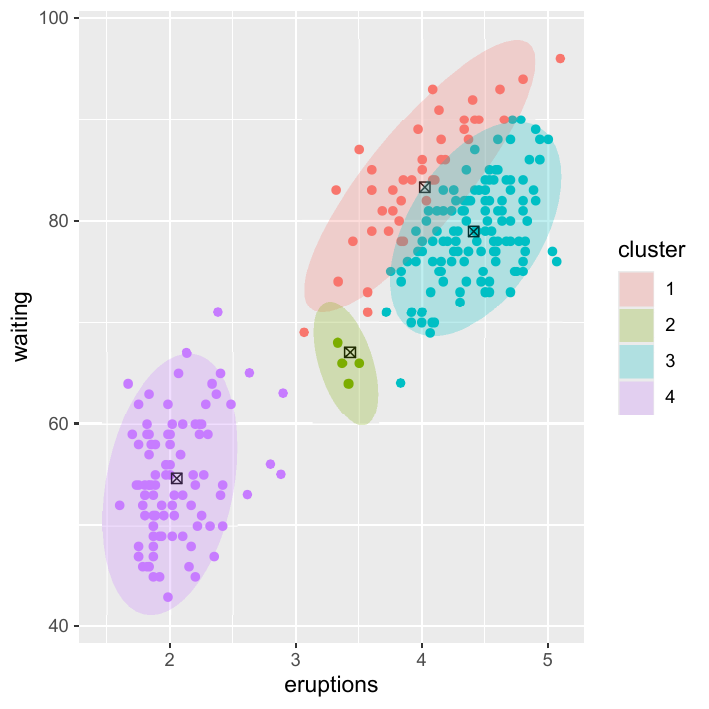}
\caption{$W$ = 0.001} \label{fig:4a}
\end{subfigure}
\begin{subfigure}{0.41\textwidth}
\includegraphics[width=\linewidth]{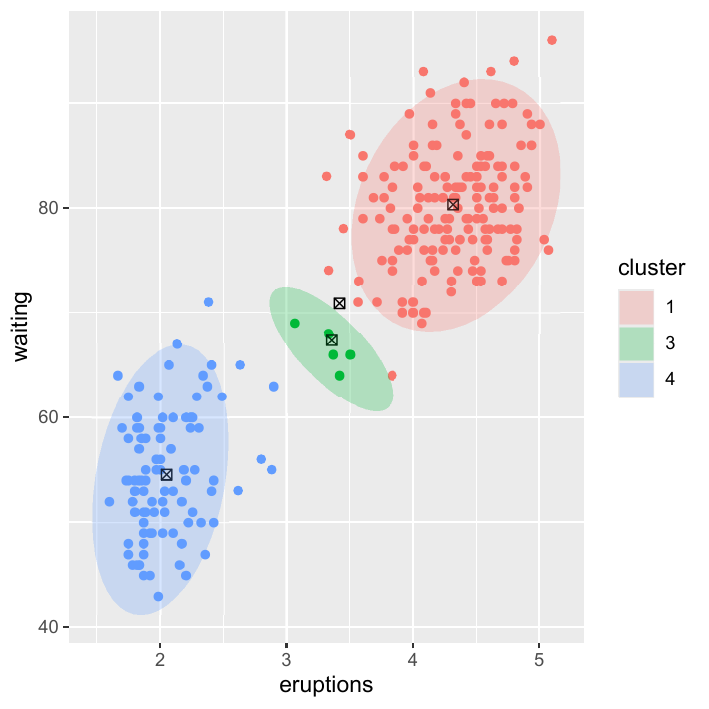}
\caption{$W$ = 2.001} \label{fig:4b}
\end{subfigure}
\caption{\label{fig:w} Posterior clustering with two different $W$ vector prior
hyperparameter settings. (a) low $W$ vectors; (b) high $W$ vectors.}
\end{figure}

The $W$ hyperparameter of the Normal–Wishart prior sets the preferred shape and scale of cluster covariance,
as illustrated in Figure~\ref{fig:w}. Low values of $W$ imply smaller prior precision, which in turn
yields more diffuse covariance estimates for the Gaussian components. This increases the
risk of overfitting by allowing the posterior covariances to adapt too freely to the data.
Conversely, higher values of $W$ imply larger prior precision and therefore smaller
component covariances, strengthening prior regularisation and potentially inducing underfitting.

\newpage
\begin{CodeChunk}
\begin{CodeInput}
R>   gen_path <- tempdir()
R>   w_grid <- data.frame(c1=c(0.001,2.001),
R>                        c2=c(0.001,2.001),
R>                        c3=c(0.001,2.001),
R>                        c4=c(0.001,2.001))
R>   init <- "kmeans"
R>   k <- 4
R>
R>   for (i in 1:nrow(w_grid)) {
R>     w0 = diag(w_grid[,i],P)
R>     prior <- list(
R>       W = w0,
R>       logW = -2*sum(log(diag(chol(w0))))
R>     )
R>
R>     gmm_result <- vb_gmm_cavi(X=X, k=k, prior=prior, delta=1e-8, init=init,
R>                               verbose=FALSE, logDiagnostics=FALSE)
R>     do_prior_plots(i, gmm_result, "w", w_grid, gen_path)
R>  }
\end{CodeInput}
\end{CodeChunk}

\clearpage

\subsection{Effect of initialiser on the Variational Gaussian Mixture Model} \label{sec:init}

The VB GMM can be initialised using one of three methods:

\begin{enumerate}
  \item kmeans
  \item dbscan
  \item random assignment
\end{enumerate}

The default initialisation method is \code{kmeans} (\cite{Lloyd:1982}), as it is simple,
computationally fast and gives reasonable estimates if the data are not too complex.
In cases of complex data, such as the Sickle Cell Disease sample data included in the
package, we recommend \code{DBSCAN} by \cite{Hahsler+Piekenbrock+Doran:2019}. \code{Random initialisation} is included for
completeness, but it is rarely useful because variational inference often becomes trapped
in local minima. If \code{DBSCAN} generates more components in the result than required, we
recommend Hierarchical \code{DBSCAN} to merge unwanted clusters.

\subsubsection{VB GMM example using kmeans}

\begin{figure}[t!]
\centering
\begin{subfigure}{0.31\textwidth}
\includegraphics[width=\linewidth]{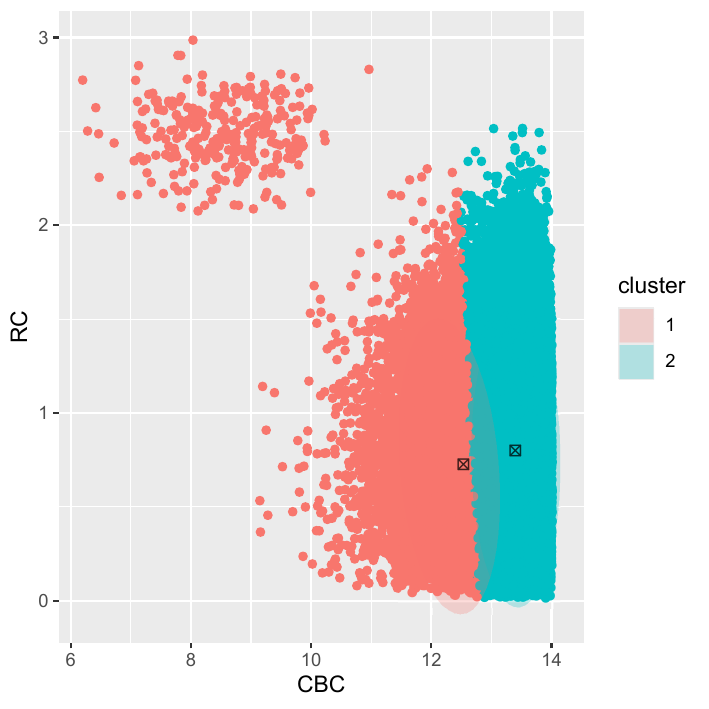}
\caption{k = 2} \label{fig:5a}
\end{subfigure}
\begin{subfigure}{0.31\textwidth}
\includegraphics[width=\linewidth]{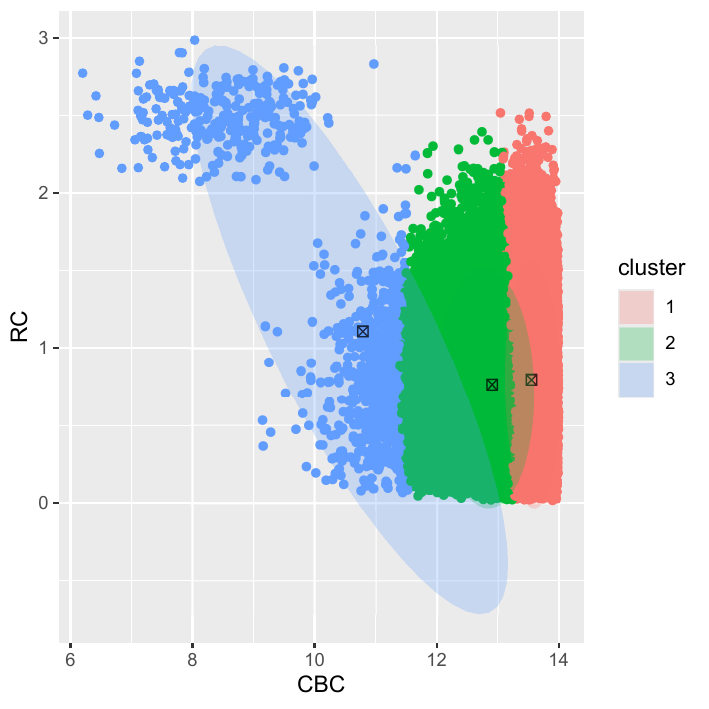}
\caption{k = 3} \label{fig:5b}
\end{subfigure}
\begin{subfigure}{0.31\textwidth}
\includegraphics[width=\linewidth]{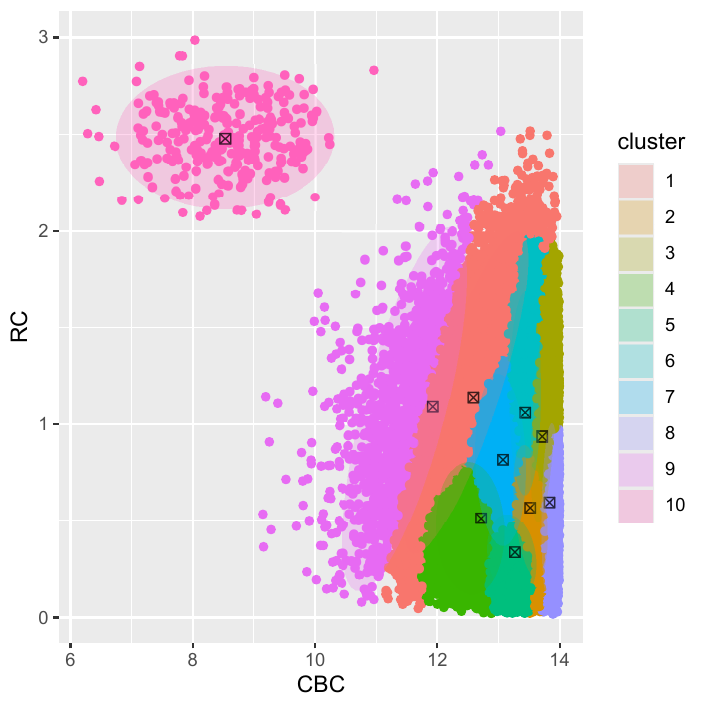}
\caption{k = 10} \label{fig:5c}
\end{subfigure}
\caption{\label{fig:kmscd} Posterior clustering of the SCD cohort using k-means initialisation.
(a) The desired k=2; (b) increasing k to 3 and (c) increasing k to 10.}
\end{figure}

Figure~\ref{fig:kmscd} illustrates the challenges when using \code{kmeans} as an initialiser
for an extremely unbalanced data set, e.g. the \code{scd_cohort} data included in the
\pkg{VBphenoR} package. With (a) k=2, the model is unable to separate the rare SCD subgroup from
the majority class. Increasing to (b) k=3 yields partial separation, but substantial overlap
remains and the inferred centroid is still far from the true SCD cluster. Only when k is increased
to (c) k=10 does the model reliably identify the SCD cluster; however, at this setting the
majority class is split into multiple subclusters.  It is clear that \code{kmeans} is unable to
cope with very noisy and highly unbalanced cluster classes.

The code for Figure~\ref{fig:kmscd} is shown below.
(note the \code{do_init_plot()} function is detailed in Appendix~\ref{app:plots}).

\begin{CodeChunk}
\begin{CodeInput}
R>  run_gmm <- function(k, init, initParams) {
R>    data(scd_cohort)
R>    x <- scd_cohort[,.(CBC,RC)]
R>    p <- ncol(x)
R>    prior <- list(
R>      alpha = 0.001,
R>      beta = 1,
R>      m = matrix(rowMeans(t(x),ncol=1),
R>      v = p+1,
R>      W = diag(1,p),
R>      logW = -2*sum(log(diag(chol(diag(1,p)))))
R>    )
R>
R>    gmm_result <- vb_gmm_cavi(X=x, k=k, prior=prior, delta=1e-6, maxiters = 5000,
R>                              init=init, initParams=initParams,
R>                              stopIfELBOReverse=TRUE, verbose=FALSE)
R>
R>    do_init_plot(scd_cohort, gmm_result)
R>  }
R>
R>  run_gmm(k = 2, init = "kmeans", initParams = NULL)
R>  run_gmm(k = 3, init = "kmeans", initParams = NULL)
R>  run_gmm(k = 10, init = "kmeans", initParams = NULL)

\end{CodeInput}
\end{CodeChunk}

\subsubsection{VB GMM example using DBSCAN}

\begin{figure}[t!]
\centering
\includegraphics[width=0.7\linewidth]{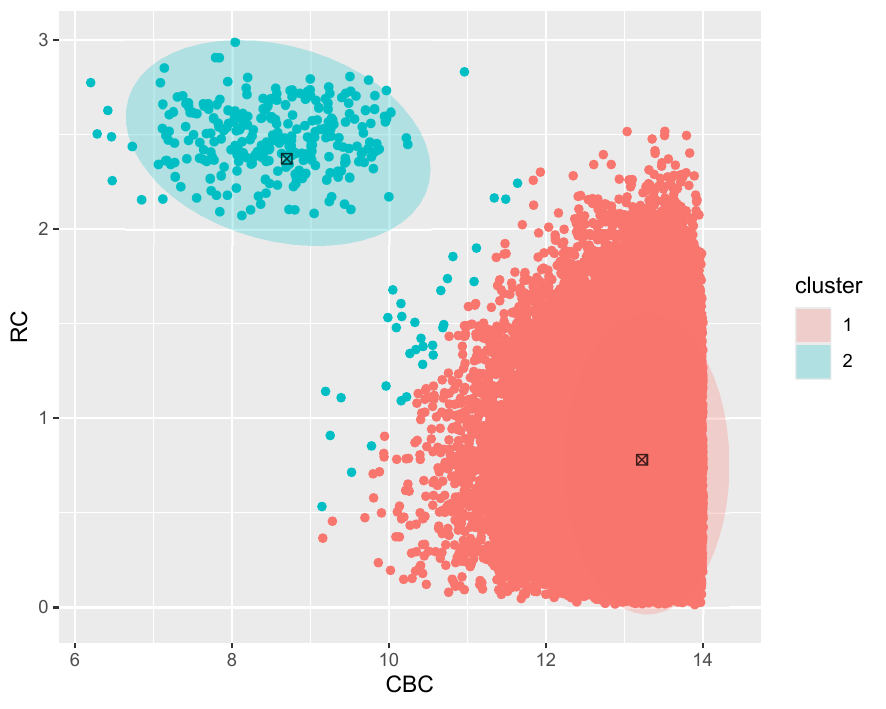}
\caption{\label{fig:dbsscd} Posterior clustering of SCD Cohort with \code{DBSCAN}
initialisation using k=2. The SCD cluster is identified by the model, albeit with some
misclassified non-SCD observations on the boundary with the majority class.}
\end{figure}

Figure~\ref{fig:dbsscd} is a single run using \code{DBSCAN} as the initialiser
for the highly unbalanced and noisy \code{scd_cohort} data.  \code{DBSCAN} is very
capable of finding the rare phenotype class and thus provides excellent starting
positions for the VB GMM. This is at the cost of computational performance.
While \code{kmeans} runs in less than a second, \code{DBSCAN} takes about 16 seconds
for these data.  Another issue with \code{DBSCAN} is that it cannot be limited to
a specified number of clusters k, so it often returns more than k clusters when k
is small e.g. k = 2 in our current phenotyping model.  We therefore need to merge
clusters returned by \code{DBSCAN} before passing the correct number k cluster
centroids to VB GMM.

The code for Figure~\ref{fig:dbsscd} is shown below.

\begin{CodeChunk}
\begin{CodeInput}

R>  initParams <- c(0.15, 5)
R>  names(initParams) <- c('eps','minPts')
R>  run_gmm(k = 2, init = "dbscan", initParams = initParams)

\end{CodeInput}
\end{CodeChunk}

\subsection{Variational Logistic Regression Model} \label{sec:Variational Logit Model}

We employ a variational logistic regression to classify binary indicators such as availability
of biomarker information and clinical codes, given the latent class we obtained from the VB GMM.
In canonical logistic regression, the response variable $y_i$ is binary and the likelihood
for $y_i$ is

\begin{equation}
P(y_i|X_i,\beta) = \sigma(X^T_i\beta)^{y_i} \cdot  [1 - \sigma(X^T_i\beta)]^{1-y_i}
\end{equation}

where

\begin{itemize}
  \item $X_i$ is the feature vector for observation $i$,
  \item $\beta$ is the vector of coefficients,
  \item $\sigma(z) = \frac{1}{1 + e^{-z}}$ is the logistic i.e. sigmoid function.
\end{itemize}

In our patient phenotyping model, $y_i$ can be the GMM soft probabilities and $\beta$ can
be the coefficients for phenotype predictors.

\subsection{Effect of informative priors on the Variational logistic regression Model}

Informative priors play an important role in the variational regression model. In
this section we briefly outline the prior effects on the VB Logit posterior
estimates.  To perform Bayesian inference, we specify a prior distribution for the
regression coefficients $\beta$. A common choice is a multivariate normal prior
for the true posterior (\cite{Gelman+Jakulin+Pittau+Su:2008}):

\begin{equation} \label{eq:logibeta}
P(\beta) = \mathcal{N}(\beta|m_0, S_0)
\end{equation}

where

\begin{itemize}
  \item $m_0$ is the prior mean vector assumption for the true (unknown) mean vector,
  \item $S_0$ is the prior covariance matrix for the true (unknown) covariance matrix.
\end{itemize}

In variational inference, the prior is part of the Kullback-Leibler divergence
(\cite{Kullback+Leibler:1951}) between the approximate posterior $Q(\beta)$ and the
true posterior $P(\beta)$:

\begin{equation} \label{eq:kl}
KL(Q(\beta) || P(\beta))
\end{equation}

where

\begin{equation} \label{eq:qbeta}
Q(\beta) = \mathcal{N}(\beta|m,S)
\end{equation}

and,

\begin{itemize}
  \item $m$ is the best-fit approximate mean vector,
  \item $S$ is the best-fit approximate covariance matrix
\end{itemize}

This effect is used for optimising the variational parameters by maximising the
Evidence Lower Bound (ELBO), which is a lower bound on the log marginal likelihood:

\begin{equation} \label{eq:elbo}
ELBO = \mathbf{E}_{Q(\beta)}[log P(y|X,\beta)] - KL(Q(\beta)||P(\beta))
\end{equation}

If we use uninformative priors for $m_0$ and $S_0$, we reduce the model to essentially
maximum likelihood given the data. In clinical settings, we usually have expert
medical opinion or empirical clinical evidence, such as patient history, that can
be used to set informative priors. This approach can enhance the clinical value of
the posterior results and overcome some of the limitations in EHR data, such as
missing biomarker data and high levels of imbalance in the responses. Our SCD example
in Section~\ref{sec:algorithm} uses informative priors for patient characteristics age
and highrisk and the two biomarkers.


\section{Summary and discussion} \label{sec:summary}

In this paper, we present \pkg{VBphenoR}, a software package for Bayesian patient phenotyping.
The package implements two complementary variational inference methods: a variational
Gaussian mixture model for estimating latent phenotypes from patient characteristics,
and variational logistic regression for incorporating additional patient history data,
such as clinical codes and medication records. The important effects of informative
priors have been examined in some detail. For the variational Gaussian mixture model,
the choice of initialisation method is shown to significantly influence result accuracy.

It is hoped to extend the package in the near future including use of \code{C} inside \code{R}
to enhance computational efficiency. Broader applicability across diverse disease
conditions could be achieved by extending the regression framework to accommodate
polytomous (multiclass) phenotype outcomes.

\section*{Acknowledgments}

The authors wish to thank the article referees for providing many helpful comments
and suggestions, both for this article and the \pkg{VBphenoR} package.


\bibliography{refs}


\newpage

\begin{appendix}

\section{Technical details} \label{app:plots}

The utility code, \code{do_prior_plots} and \code{do_init_plot}, used to create the plots in the paper:

\begin{CodeChunk}
\begin{CodeInput}
R>   do_prior_plots <- function(i, gmm_result, var_name, grid, fig_path) {
R>     dd <- as.data.frame(cbind(X, cluster = gmm_result$z_post))
R>     dd$cluster <- as.factor(dd$cluster)
R>
R>     mu <- as.data.frame( t(gmm_result$q_post$m) )
R>     cols <- c("#1170AA", "#55AD89", "#EF6F6A", "#D3A333", "#5FEFE8", "#11F444")
R>     p <- ggplot2::ggplot() +
R>       ggplot2::geom_point(dd, mapping=ggplot2::aes(x=eruptions,
R>                                                    y=waiting,
R>                                                    color=cluster)) +
R>       ggplot2::scale_color_discrete(cols, guide = 'none') +
R>       ggplot2::geom_point(mu, mapping=ggplot2::aes(x = eruptions, y = waiting),
R>                           color="black",
R>                           pch=7, size=2) +
R>       ggplot2::stat_ellipse(dd, geom="polygon",
R>                   mapping=ggplot2::aes(x=eruptions, y=waiting, fill=cluster),
R>                    alpha=0.25)
R>
R>     grids <- paste((grid[,i]), collapse = "_")
R>     ggplot2::ggsave(filename=paste0(var_name,"_",grids,".png"), plot=p,
R>                     path=fig_path,
R>                     width=12, height=12, units="cm", dpi=600,
R>                     create.dir = TRUE, device=cairo_ps)
R>   }
R>
R>
R>   do_init_plot <- function(scd_cohort, gmm_result) {
R>     dd <- as.data.frame(cbind(scd_cohort, cluster=gmm_result$z_post))
R>     dd$cluster <- as.factor(dd$cluster)
R>     dd$highrisk <- as.factor(dd$highrisk)
R>     mu <- as.data.frame( t(gmm_result$q_post$m) )
R>
R>     cols <- c("#1170AA", "#55AD89", "#EF6F6A", "#D3A333", "#5FEFE8", "#11F444")
R>     p <- ggplot() +
R>       geom_point(dd, mapping=aes(x=CBC, y=RC, color=cluster)) +
R>       scale_color_discrete(cols, guide = 'none') +
R>       geom_point(mu, mapping=aes(x = CBC, y = RC), color="black", pch=7, size=2) +
R>       stat_ellipse(dd, geom="polygon",
R>                    mapping=aes(x=CBC, y=RC, fill=cluster),
R>                    alpha=0.25)
R>
R>     return(p)
R>   }
\end{CodeInput}
\end{CodeChunk}
\newpage

\end{appendix}


\end{document}